# Discharge characteristics and parameter diagnosis of dielectric barrier discharge patterns in double-gap configuration[∗]

TIAN Shuang[1], ZHANG Han[1], ZHANG Xi[1], ZHANG Xuexue[1], LI Xuechen[1], LI Qing[1,2], RAN Junxia[1,2]

1.College of Physics Science and Technology, Hebei University, Baoding 071002, China

2.Engineering Research Center of Zero-carbon Energy Buildings and Measurement Techniques, Ministry of Education, Baoding 071002, China

**Abstract**

Pattern discharge is a common mode in dielectric barrier discharge (DBD) and has broad application prospects in various industrial fields, such as material surface treatment, environmental monitoring, and biomedical applications. In this work, a mixed gas of 75% argon and 25% air is used to generate a pattern discharge. A double-gap boundary composed of hexagonal configuration and square configuration is employed, and the gas pressure is fixed at 20 kPa. By varying the applied voltage amplitude, single-ring pattern, square-point-line pattern, square lattice pattern, and annular-lattice pattern are obtained for the first time. The discharge characteristics and their temporal correlation are studied using both optical method and electrical method. The results show that the discharge patterns exhibit multiple discharges in each half of the voltage cycle, and these discharges are temporally correlated with each other. Time-resolved discharge images of the square lattice pattern are captured using an enhanced charge-coupled device (ICCD). The experimental results reveal that multiple discharges in a half-voltage cycle correspond to the ignition process of the pattern in the radial direction from the outside to the inside. The morphology of the square lattice pattern observed by the naked eye is actually the result of the temporal superposition of luminescence from points at different positions in the evolution process. The formation mechanism of this pattern is analyzed through electric field simulations and theoretical calculations. Plasma parameters are diagnosed by collecting the emission spectrum of the square dot-lattice pattern. The results show that the electron density gradually decreases radially from the outer region to the inner region, while the electron temperature and molecular vibrational temperature increase radially from the outer region to the inner region, and the molecular rotational temperature remains almost unchanged. The temporal evolution of the square lattice pattern is shown in the following figures, where the current waveform marks the timing of each frame of ICCD imaging for the complete square lattice pattern.

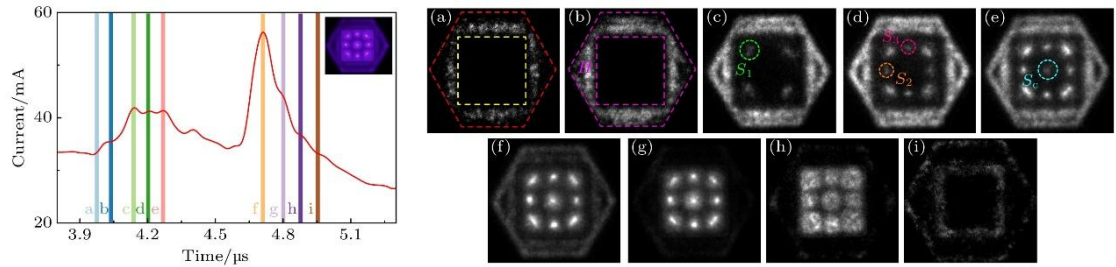

PACS：52.35.Mw, 47.54.–r, 52.80.Tn

**doi:** 10.7498/aps.74.20250111

**cstr:** 32037.14.aps.74.20250111

# 1. Introduction

Dielectric barrier discharge (DBD) is considered to be one of the economical, widespread and reliable methods for generating low temperature plasma[1–3]. It is widely used in various technical fields such as surface treatment[4], ozone generation[5,6], catalytic[7,8], biomedical[9,10] and material deposition[11]. Generally speaking, DBD can show three modes in space: filamentary discharge mode formed by a large number of randomly distributed discharge filaments, pattern mode composed of microdischarges with spatial and temporal distribution, and uniform or diffuse mode[12,13]. Under certain conditions, it is easier to achieve uniform or diffuse discharge by using helium or neon as the working gas[14–16]. When the working gas is argon or a mixture of gases, the filamentary discharge mode is the most common[17,18]. Under suitable conditions, the microdischarge channels between the electrodes can self-organize to form patterns with fine and orderly macrostructures[19,20]. Patterns formed in DBD systems have attracted much attention in the past 20 years[21]. The experimental conditions such as voltage amplitude[22], driving frequency[23], gap width[24], gas content[25], gas pressure[26] and so on can affect the structure of the pattern. In addition, if there is a boundary in the discharge region, the electric field near the boundary will be distorted, which makes the distribution of the electric field and wall charge near the boundary different from that inside the boundary, thus affecting the formation and evolution of the self-organized pattern[27–29]. By changing the experimental conditions, the researchers observed a number of spatially regular and stable patterns, including square pattern[30,31], honeycomb hexagonal pattern[32], stripe pattern[33], concentric ring pattern[34] and superlattice pattern[35]. It is found that the regular pattern in space also shows a regular development process in time[36,37]. Pan et al.[38] discovered a honeycomb-Kagome hexagonal superlattice pattern with dark discharge. Its spatiotemporal structure is composed of three interlaced sublattices. There are three current pulses in each half voltage cycle, and each current pulse corresponds to a set of sublattices. Li et al.[39] found a three-dimensional pattern composed of five sublattices. There are three current pulses in each half voltage cycle, and some of the current pulses contain two sublattice discharges. The spatiotemporal structure of the pattern is the result of the overlapping of sublattices. This shows that the evolution process of patterns is diverse, and the difference of the evolution process is related to the formation mechanism. Therefore, the mechanism can be studied in depth through the analysis of the evolution process.

In this paper, a special combined air gap is designed, and a series of new patterns are observed under the double boundary condition composed of regular hexagon and square. The time evolution behavior of the square lattice pattern is studied by using the short exposure photographs taken by the enhanced charge coupled device (ICCD), and the plasma parameters of the square lattice pattern were diagnosed by using the emission spectrum. The pattern enriches the diversity of patterns in DBD, and the related research provides a theoretical reference for the study of patterns in other experimental systems.

## 2. Experimental setup

The schematic diagram of the experimental setup is shown in Fig. 1(a). Two cylindrical containers with an inner diameter of 80 mm are placed opposite to each other, and both ends are sealed with a quartz glass plate with a thickness of 1 mm as the dielectric layer of the DBD. The container is filled with water, and the two copper rings are immersed in the water respectively to form the water electrode of the DBD. One copper ring is connected to a sinusoidal AC power supply (Nanjing Suman CTP-2000 K), and the other copper ring is grounded. Two quartz glass frames with a square hole (side length of 20 mm) and a regular hexagonal hole (side length of 20 mm) with a thickness of 2 mm are fixed between the two water electrodes and pressed. The arrangement of the dielectric layer and the quartz glass frame is shown in the Fig. 1(a). During the experiment, the two quartz glass frames are closely attached and fixed between the two quartz dielectric layers and pressed, as shown in the side view of the Fig. 1(b). That is, the air gap distance in the quadrilateral area in the figure is 4 mm, and the air gap distance in the area enclosed by the quadrilateral boundary and the hexagonal boundary is 2 mm. The whole apparatus is placed in a vacuum chamber filled with a gas mixture (argon 75%, air 25%, pressure 20 kPa). The applied voltage is measured by a high voltage probe (Tektronix P6015A) and the discharge current is measured by a current probe (Tektronix TCPA 300). The luminescence signal of the discharge is measured by a photomuliplier tube (PMT) (ET 9130/100B) after passing through a quartz lens. The waveforms of the applied voltage, the discharge current, and the luminescence signal are recorded and displayed by an oscilloscope (Tektronix DPO4104). Discharge images are taken with a digital camera (Canon EOS 5D) and an intensified charge-coupled device (ICCD) (Andor DH334T). A self-made trigger is used to convert the sinusoidal signal exciting the DBD into a synchronous transistor-transistor logic (TTL) signal, which is used to trigger the ICCD and oscilloscope. Because there is a certain jitter between the discharge time and the excitation voltage, in order to achieve the precise synchronization of ICCD shutter and discharge, the TTL and discharge luminescence signals are input to the oscilloscope at the same time, and the exposure time of ICCD relative to the discharge can be obtained on the oscilloscope by adjusting the delay time of ICCD. The luminous signal of the discharge is converged by a lens and transmitted to a spectrometer (PI ACTON SP2750) through an optical fiber (PI LG-455-020-1) to collect the emission spectrum of the discharge.

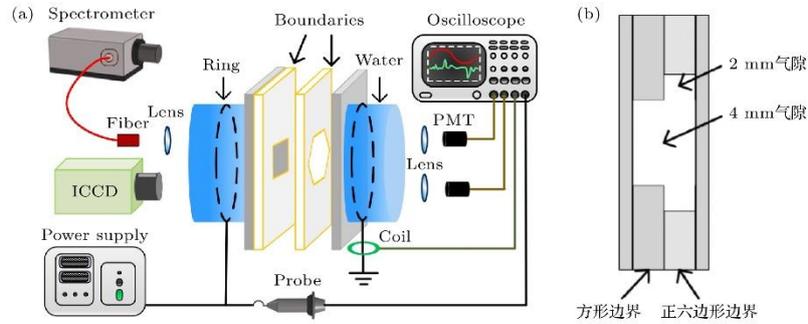

**Figure 1.** (a) Schematic diagram of the experimental setup; (b) side view of the gap structure.

## 3. Results and Discussion

The Fig. 2(a) —(d) shows the discharge pattern as a function of applied voltage. When the peak applied voltage $V_a$ is 2. 9 kV, a single ring pattern is formed, as shown in Fig. 2(a), where the halo appearing in the region enclosed by the regular hexagonal boundary to the square boundary at the 2 mm air gap is represented by $H$, the inner part of the ring region appearing in the square boundary of the 4 mm air gap is represented by $R$, and the halo outside the ring is represented by $H'$. When the $V_a$ are increased to 3.1 kV, the annular region is split into a square dot-line structure as shown in Fig. 2(b), which is called a square dot-line pattern. The luminous point near the vertex of the square boundary is represented by $S$, the luminous line near the longitudinal boundary line is represented by $L_1$, and the luminous line near the transverse boundary line is represented by $L_2$. With the further increase of $V_a$ and to 3.3 kV, the luminous line in the square dot-line pattern becomes a luminous point, and a luminous point appears in the center area, which is called the square lattice pattern, as shown in Fig. 2(c). In the square lattice pattern, the points near the vertex of the square boundary are represented by and $S_1$, the points near the longitudinal boundary are represented by $S_2$, the points close to the transverse boundary are represented by $S_3$, and the points at the center of the square boundary are denoted by $S_c$. When the $V_a$ continue to rise to 3.5 kV, the $S_2$, $S_3$ and $S_c$ in the square lattice pattern split to form a annular-lattice pattern , as shown in Fig. 2(d). At this time, the points near the vertices of the square boundary in the rectangular lattice pattern are represented by $S'_1$, the points near the longitudinal boundary are represented by $S'_2$, the points close to the transverse boundary are denoted by $S'_3$, and the point at the center of the rectangular lattice is denoted by $S'_4$.

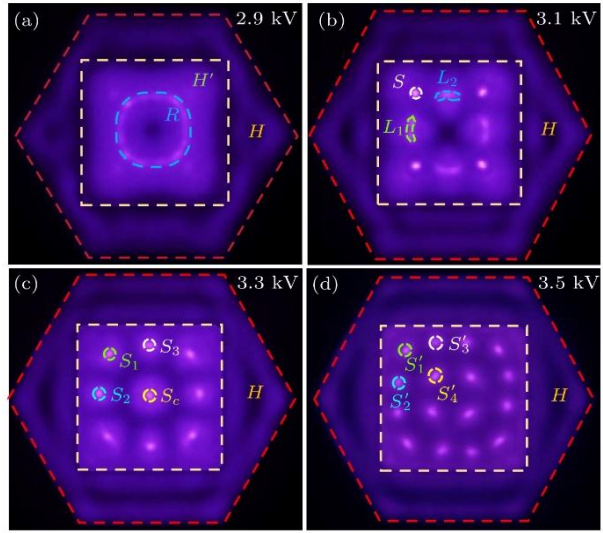

**Figure 2.** Patterned discharges under different applied voltages with an exposure time of 0.1 s.

Fig. 3(a) —(d) is Fig. 2(a) — waveform of applied voltage and discharge current of discharge corresponding to(d). It is found that there are always two current pulses in each half voltage cycle in the single ring pattern, the square dot-line pattern and the square lattice pattern. The intensity of the first current pulse is small and the intensity of the second current pulse is large. For the annular-lattice pattern, there are four different current pulses in each half voltage cycle, and the amplitude of the first current pulse is the largest.

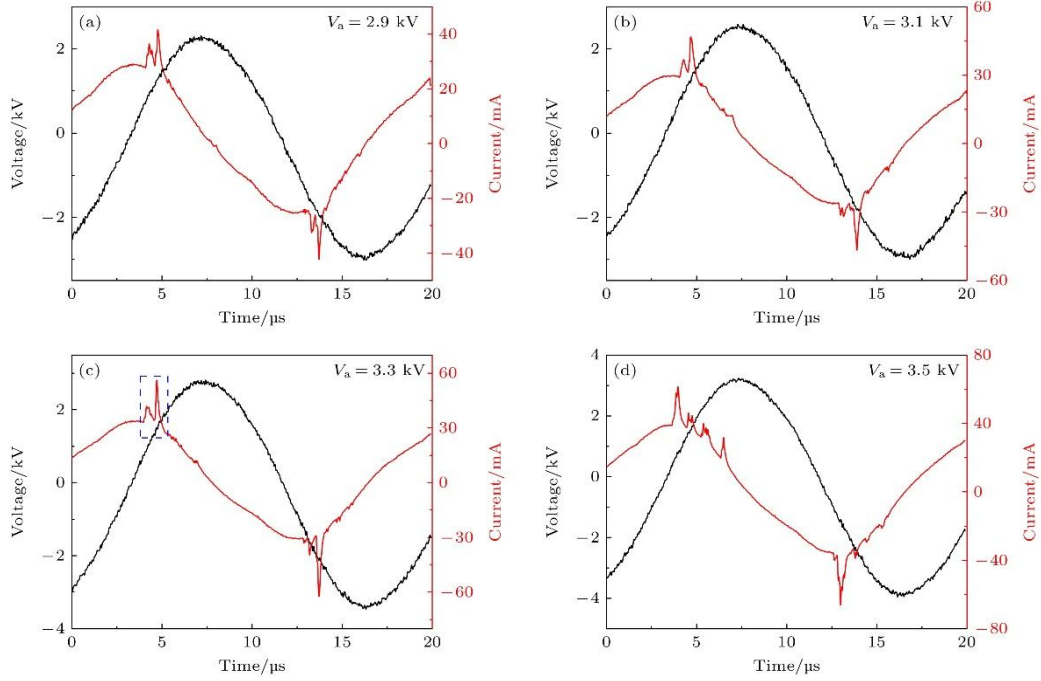

**Figure 3.** Waveforms of applied voltage and discharge current for different patterns.

In order to study the discharge time behavior of each luminous point of the pattern, the *H* region in the Fig. 2 has a small overall change, so the overall study is selected, and the rest part of the pattern is selected to study according to the symmetry of the pattern. The optical properties of the discharge filament at different positions marked by four kinds of patterns in the Fig. 2 are measured accurately by using a highly sensitive photomultiplier tube, as shown in the Fig. 4. The results show that the region enclosed by the regular hexagonal boundary to the square boundary (*H*) discharges first in the four patterns. For a single ring pattern, a ring (*R*) within a 4 mm air gap immediately begins to discharge, followed by a corona (*H'*), as shown in Fig. 4(a). For the square dot-line pattern, the luminous points (*S*) near the four vertices of the square boundary in the 4 mm air gap discharge first, followed by the luminous lines ($L_1$ and $L_2$) near the boundary line, as shown in Fig. 4(b). For the square lattice pattern, in the 4 mm air gap, the luminous points near the four vertices of the square boundary ($S_1$) discharge first, then the luminous points near the boundary line ($S_2$ and $S_3$) discharge, and finally the luminous points in the center region ($S_c$) discharge, as shown in Fig. 4(c). For the annular-lattice pattern, the luminous points ($S'_1$) near the four vertices of the square boundary in the 4 mm air gap discharge first, then the luminous points ($S'_2$ and $S'_3$) near the boundary line discharge, and finally the luminous point ($S'_4$) in the center of the rectangle discharges. In the 4 mm air gap, it can be clearly observed that the closer to the boundary, the earlier the discharge time, that is, the corresponding discharge shows an overall development trend from outside to inside.

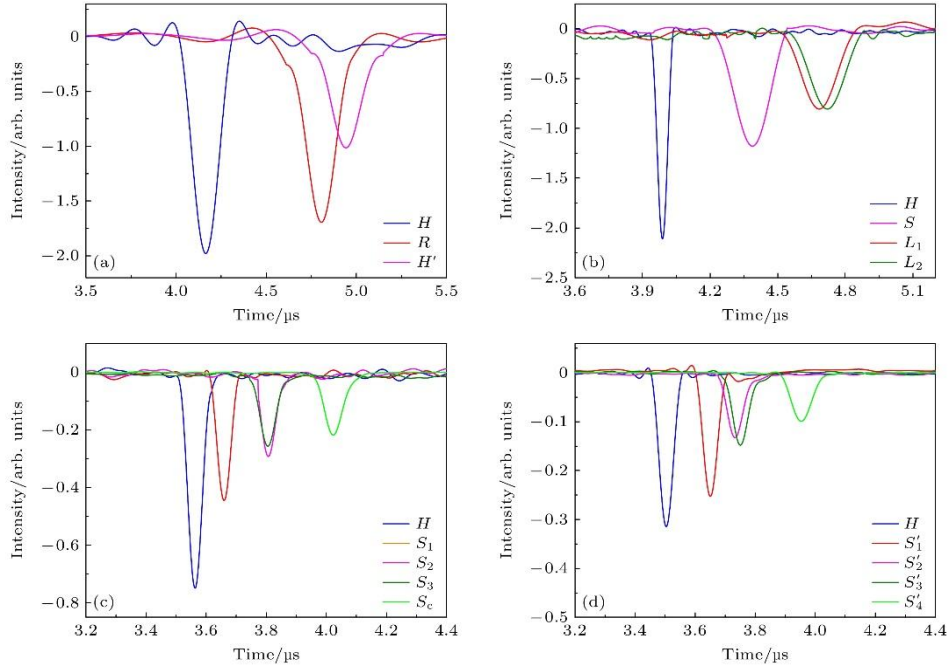

**Figure 4.** Temporal correlation from the pattern: (a)–(d) Correspond to Fig. 2(a)-(d), respectively.

In order to study the time evolution behavior of the whole discharge pattern more clearly, the square lattice pattern is photographed by ICCD camera with short exposure time. The waveform above the Fig. 5 is a partial magnification of the current pulse enclosed by the blue line in the Fig. 3(c). Below the Fig. 5 is a single discharge image taken during the discharge (exposure time $t_{exp}$ = 20 ns), where the shooting times of Fig. 5(a) —(i) correspond to the times a — i on the waveform above Fig. 5, respectively. From the Fig. 5, it can be found that the discharge first occurs in the 2 mm air gap region (time a). With the increase of the discharge current, the discharge gradually fills the whole 2 mm air gap region (time b). Then, a discharge filament appears at the position of $S_1$ in the region near the vertex of the square boundary (time c). Then the filament appears at $S_2$ and $S_3$ (time d), and finally the filament appears at $S_c$ (time e). At the time of current maximum (time f), the complete square lattice pattern can be clearly observed. Then, the discharge begins to weaken gradually, first of all, the discharge in the regular hexagonal boundary region begins to weaken (time g). After that, the discharge in the square boundary region begins to weaken as a whole (time h). At time i, there is still discharge near the edges of the two boundaries, and then the discharge gradually disappears with the decrease of the current. In the above process, it can be found that the square lattice pattern is gradually formed from the outside to the inside along the radial direction, which is consistent with the previous result of measuring the luminous signal at each position by using a photomultiplier tube, that is, the morphology of the square lattice pattern seen by the naked eye is actually the superposition of these points at different positions in the evolution process shown in Fig. 5.

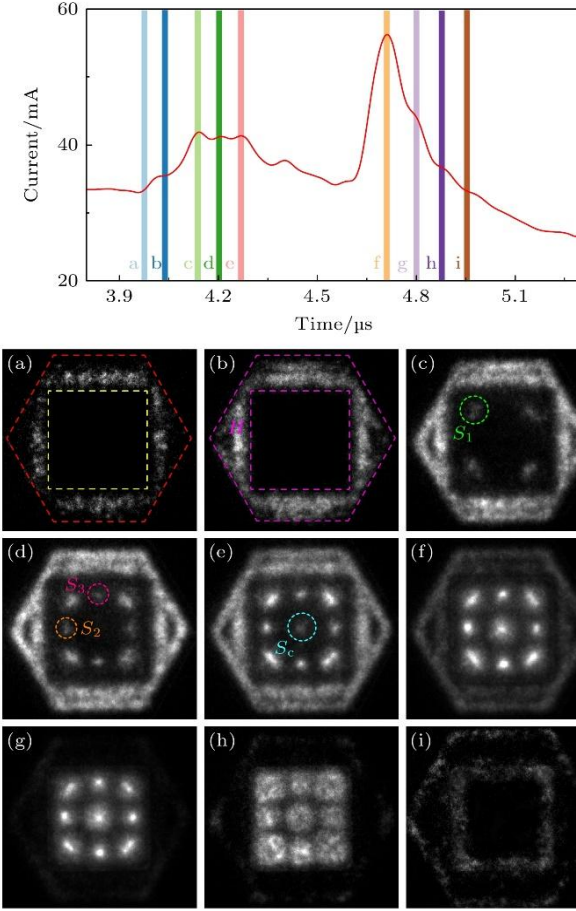

**Figure 5.** Temporal evolution of the square lattice pattern. The current waveform shows the ICCD shooting time of each picture, and the exposure time is 20 ns (single shot).

When the applied voltage is the same, the electric field in different air gaps is different. The breakdown voltage of different air gaps can be calculated by the equivalent circuit, and then the breakdown electric field can be obtained[40]. The Fig. 6(a) is the equivalent circuit of the dielectric barrier discharge. Before the first breakdown, the equivalent capacitance of the air gap $C_g$ and the equivalent capacitance of the barrier dielectric $C_{m1}$, $C_{m2}$ are in series, and they divide the applied voltage $V_a(t)$. The increase of air gap voltage $V_g(t)$ completely depends on the increase of $V_a(t)$. Only when $V_a(t)$ reaches a certain value, which makes $V_g(t)$ reach the static breakdown voltage $V_{b0}$ of the air gap, can the air gap breakdown occur. $V_{b0}$ can be calculated from (1). In the experiment, the thickness of the barrier medium is different for different air gaps, so different $C_g$ correspond to different $C_{m1}, C_{m2}$. $C_{m1}, C_{m2}$ can be calculated from the plate capacitance formula. $C_m$ is the sum of the equivalent capacitance of the barrier medium, calculated by (2). The breakdown electric field $E$ can be obtained from (3):

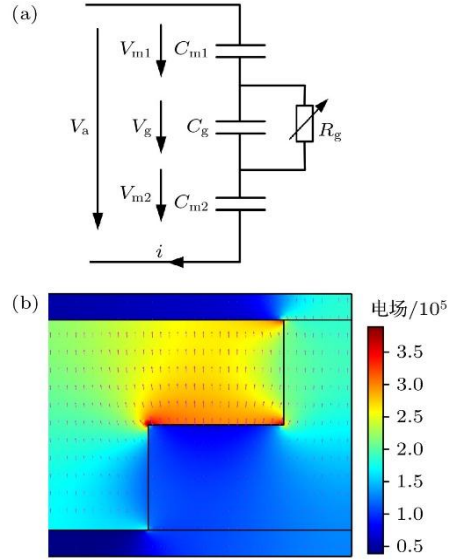

**Figure 6.** (a) Equivalent circuit diagram of dielectric barrier discharge; (b) simulation results of applied electric field.

$$V_{b0} = \frac{C_m}{C_m + C_g} V_a, \tag{1}$$

$$\frac{1}{C_m} = \frac{1}{C_{m1}} + \frac{1}{C_{m2}}, \tag{2}$$

$$E = \frac{V_{b0}}{d}. \tag{3}$$

According to the formula (1) — (3), the electric field at 2 mm air gap is 480 kV/m, which is higher than the electric field at 4 mm air gap 330 kV/m. According to the theory of gas discharge, the place with high electric field intensity discharges first, so the regular hexagonal boundary region ($H$) discharges first, which is also consistent with the simulation results of electrostatic field in Fig. 6(b). From the Fig. 6(b), it can be seen that the electric filed at 2 mm air gap is stronger than that at 4 mm air gap. In the following process, as the applied voltage continues to rise, the discharge develops to the transition region between the 2 mm air gap and the 4 mm air gap, as shown in the Fig. 6(b), the electric field near the boundary is the strongest, and the electric field at the intersection of the two boundaries, that is, the vertex of the quadrilateral, will be stronger than that in other surrounding regions. Therefore, the discharge filament will be generated earlier at the position near the vertex angle of the boundary ($S_1$). Gas discharge is affected by many factors, such as electric field uniformity, electrode shape, gas properties and so on. When the applied electric field strength exceeds the gas breakdown threshold, the gas is broken down and micro-discharges are formed, and due to the conversion of the polarity of the power supply, the micro-discharges will appear repeatedly at the same position, forming a micro-discharge channel. Positive and negative charged particles and particles with excited state will be produced at the same time of the formation of micro-discharge channels. Under the action of an applied electric field, these particles will move to the positive and negative electrodes and eventually deposit on the dielectric slab to form wall charges. When a large number of wall charges are deposited on the dielectric surface, a new electric field is formed, which is called wall charge field. The wall charge field is opposite to the applied electric field, so the wall charge field extinguishes the discharge at that location, and the wall charge field also suppresses the discharge around the discharge channel. When a discharge filament is generated at $S_1$ and, a wall charge field is formed at this position, which inhibits the discharge around the discharge filament, so the subsequent discharge will occur at the position where the wall charge field inhibition is the weakest. Since $S_2$ and $S_3$ are located close to the boundary and are most weakly suppressed by the wall charge field at $S_1$, the discharge filaments located at $S_2$ and $S_3$ will appear after the discharge filaments are generated at $S_1$. The subsequent discharge will be affected by the three wall charge fields, and $S_c$ is the geometric center of the whole boundary, which is the weakest inhibited by the surrounding wall charge field, so the discharge filament will be generated at $S_c$. This is why the discharge occurs from the outside to the inside in the radial direction.

The Fig. 7(a) shows the emission spectrum of the square lattice pattern in the range of 300 — 800 nm. Many molecular bands and atomic spectral lines were observed from the distribution of the spectral lines. The typical spectral lines include the $N_2$ of the second positive band system of $N_2$ ($C^3\Pi_u \rightarrow B^3\Pi_g$), the nitrogen molecular ion spectral line $N_2^+$ (B-X) and the emission line of Ar I (4p → 4s). The plasma parameters of each microdischarge channel were estimated from the above spectral lines. The molecular rotational temperature is calculated by fitting the nitrogen molecular ion spectral line $N_2^+$ (B-X) at 391.4 nm. As shown in Fig. 7(b), the gas temperature of the non-equilibrium plasma ($T_g$) is approximately equal to its molecular rotational temperature ($T_{rot}$)[41]. It is found that the molecular rotational temperatures of various microdischarge channels are basically the same, about (400 ± 10) K; The molecular vibrational temperature is calculated by Boltzmann fitting[42] using the nitrogen molecular spectral line of $N_2$ ($C^3\Pi_u \rightarrow B^3\Pi_g$), as shown in Fig. 7(c). It can be seen that the molecular vibrational temperature increases gradually from the outside to the inside along the radial direction. According to the collision-radiation model, the intensity ratios of the lines 738 nm/750 nm and 750 nm/751 nm can be used to characterize the electron density ($n_e$) and electron temperature ($T_e$), respectively[43]. Fig. 7(d) shows the change trend of $n_e$ and $T_e$ at each position of the square lattice pattern. It can be seen from Fig. 7(d) that $n_e$ gradually decreases from outside to inside along the radial direction, and $T_e$ gradually increases from outside to inside along the radial direction.

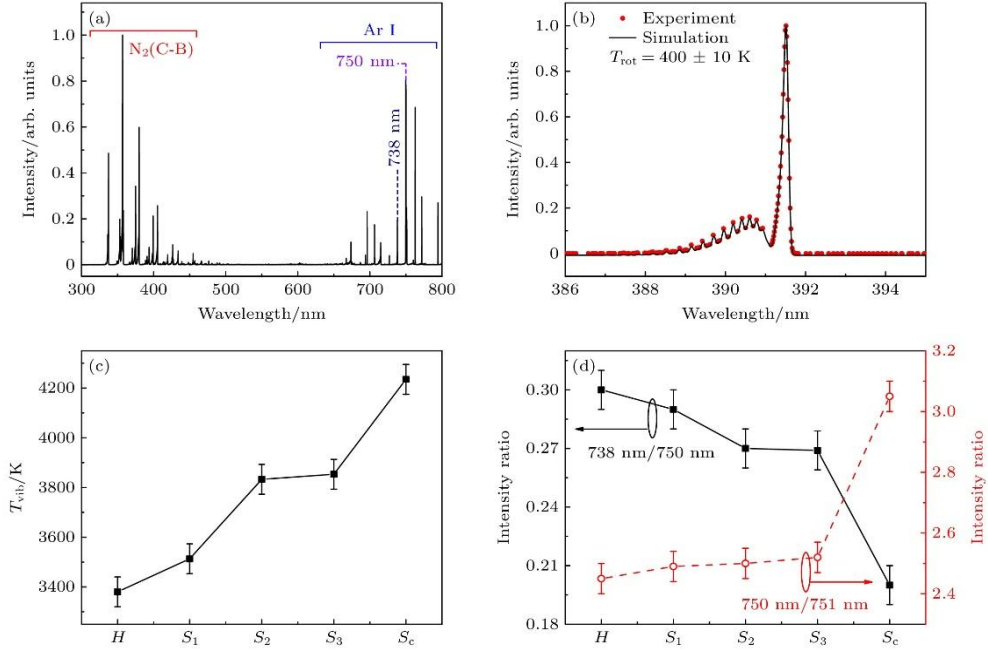

**Figure 7.** (a) 300–800 nm optical emission spectrum from the discharge; (b) fitting result of rotational bands of the $N_2^+$ (B-X); (c) $T_{vib}$ at different positions in the square lattice pattern; (d) $T_e$ and $n_e$ at different positions in square lattice pattern.

## 4. Conclusion

Using a combined boundary composed of a regular hexagon and a square, several complex patterns re generated under the excitation of a 55 kHz AC power supply, and the most representative square pattern is studied. The results show that the square lattice pattern presents two current pulses with different intensities in each half voltage cycle, and the discharge filaments forming the square lattice pattern are actually the gradual breakdown process of the discharge along the radial direction from the outside to the inside. The time evolution process of the square lattice pattern is photographed by ICCD, and its formation mechanism is analyzed theoretically. It is found that the formation process is the result of the combined action of the external electric field and the wall charge field. The diagnostic results of plasma parameters by square lattice pattern show that the gas rotational temperature is (400 ± 10) K, the molecular vibrational temperature and electron temperature increase from the outside to the inside along the radial direction, and the electron density decreases from the outside to the inside along the radial direction. These results enrich the spatiotemporal structure of DBD patterns and promote the study of pattern discharge characteristics in dielectric barrier discharge.

# References

[1] Zhang J, Tang W W, Wang Y H, Wang D Z 2023 Plasma Sources Sci. Technol. 32 055005

[2] Li J Y, Zhou D S, Rebrov E, Tang X, Kim M 2024 J. Phys. D: Appl. Phys. 57 395201

[3] Fang J L, Zhang Y R, Lu C Z, Gu L L, Xu S F, Guo Y, Shi J J 2024 Chin. Phys. B 33 015201

[4] Guan H L, Chen X R, Jiang T, Du H, Paramane A, Zhou H 2020 Chin. Phys. B 29 075204

[5] Li X C, Liu R J, Li X N, Gao K, Wu J C, Gong D D, Jia P Y 2019 Phys. Plasmas 26 023510

[6] Remy A, Geyter N D, Reniers F 2023 Plasma Process. Polym. 20 2200201

[7] Hosseini H 2023 RSC Adv. 13 28211

[8] Li Y W, Yuan H, Zhou X F, Liang J P, Liu Y Y, Chang D L, Yang, D Z 2022 Catalysts 12 203

[9] Song H, Dang Y M, Ki S H, Park S, Ha J H 2024 LWT 207 116637

[10] Domonkos M, Tichá P, Trejbal J, Demo P 2021 Appl. Sci. 11 4809

[11] Massines F, Sarra-Bournet C, Fanelli F, Naudé N, Gherardi N 2012 Plasma Process. Polym. 9 1041

[12] Akishev Y, Alekseeva T, Karalnik V, Petryakov A 2022 Plasma Sources Sci. Technol. 31 084001

[13] Tschiersch R, Nemschokmichal S, Bogaczyk M, Meichsner J 2017 J. Phys. D: Appl. Phys. 50 415206

[14] Liu K, Fang Z, Dai D 2023 Acta Phys. Sin. 72 135201

[15] Ran J X, Zhang X X, Zhang Y, et al. 2023 Plasma Sci. Technol. 25 055403

[16] Lu X P, Fang Z, Dai D, Shao T, Liu F, Zhang C, Liu D W, Nie L L, Jiang C Q 2023 High Volt. 8 1132

[17] Brandenburg R 2018 Plasma Sources Sci. Technol. 26 053001

[18] Wang Y Y, Yan H J, Li T, Bai X D, Wang X, Song J, Zhang Q Z 2023 AIP Adv. 13 085327

[19] Wang H R, Hao Y P, Fang Q, Su H W, Yang L, Li L C 2020 Acta Phys. Sin. 69 145203

[20] Wu K Y, Wu J C, Jia B Y, Ren C H, Kang P C, Jia P Y, Li X C 2020 Phys. Plasmas 27 082308

[21] Ouyang J T, Li B, He F, Dai D 2018 Plasma Sci. Technol 20 103002

[22] Hao Y P, Han Y Y, Huang Z M, Yang L, Dai D, Li L C 2018 Phys. Plasmas 25 013516

[23] Li X C, Liu R, Jia P Y, Wu K Y, Ren C H, Yin Z Q 2018 Phys. Plasmas 25 013512

[24] Li Z Y, Jin S H, Xian Y B, Nie L L, Liu D W, Lu X P 2021 Plasma Sources Sci. Technol. 30 065026

[25] Zhang Y H, Ning W J, Dai D, Wang Q 2019 Plasma Sources Sci. Technol. 28 075003

[26] Zhang J H, Pan Y Y, Feng J Y, He Y N, Chu J H, Dong L F 2023 Plasma Sci. Technol. 25 025406

[27] Duan X X, Ouyang J T, Zhao X F, He F 2009 Phys. Rev. E 80 016202

[28] Wang X, Yan H J, Wang Y Y, Yu S Q, Li T, Song J 2023 J. Phys. D: Appl. Phys. 56 105201

[29] Mokrov M S, Raizer Y P 2018 Plasma Sources Sci. Technol 27 065008

[30] Hao F, Dong L F, Du T, et al. 2018 Phys. Plasmas 25 033510

[31] Han R, Dong L F, Huang J Y, Sun H Y, Liu B B, Mi Y L 2019 Chin. Phys. B 28 075204

[32] Sun H Y, Dong L F, Liu F C, Mi Y L, Han R, Huang J Y, Liu B B, Hao F, Pan Y Y 2018 Phys. Plasmas 25 113507

[33] Wei L Y, Dong L F, Fan W L, et al. 2018 Sci. Rep. 8 3835

[34] Feng J Y, Dong L F, Wei L Y, Fan W L, Li C X, Pan Y Y 2016 Phys. Plasmas 23 093502

[35] Pan Y Y, Li Y H, Dou Y Y, Fu G S, Dong L F 2022 Phys. Plasmas 29 053502

[36] Liu F C, Liu Y N, Liu Q, et al. 2022 Plasma Sources Sci. Technol 31 025015

[37] Li C X, Dong L F, Feng J Y, Huang Y P 2019 Phys. Plasmas 26 023505

[38] Pan Y Y, Feng J Y, Li C X, Dong L F 2022 Plasma Sci. Technol. 24 115401

[39] Li Y H, Pan Y Y, Tian M, Wang Y, He Y N, Zhang J H, Chu J H, Dong L F 2023 Phys. Plasmas 30 033502

[40] Liu S H, Neiger M 2003 J. Phys. D: Appl. Phys. 36 3144

[41] Bruggeman P J, Sadeghi N, Schram D C, Linss V 2014 Plasma Sources Sci. Technol. 23 023001

[42] Yang F X, Mu Z X, Zhang J L 2016 Plasma Sci. Technol. 18 79

[43] Zhu X M, Pu Y K 2008 Plasma Sources Sci. Technol. 17 024002